\documentstyle[12pt,psfig]{article}

\begin{document}

\baselineskip=16pt plus 1pt minus 1pt

\begin{center}{\large \bf Sequence of potentials interpolating between
the U(5) and E(5) symmetries} 

\bigskip\bigskip

{Dennis Bonatsos$^{\#}$\footnote{e-mail: bonat@inp.demokritos.gr},
D. Lenis$^{\#}$\footnote{e-mail: lenis@inp.demokritos.gr}, 
N. Minkov$^\dagger$\footnote{e-mail: nminkov@inrne.bas.bg}, 
P. P. Raychev$^\dagger$\footnote{e-mail: raychev@phys.uni-sofia.bg, 
raychev@inrne.bas.bg},
P. A. Terziev$^\dagger$\footnote{e-mail: terziev@inrne.bas.bg} }

\bigskip

{$^{\#}$ Institute of Nuclear Physics, N.C.S.R.
``Demokritos''}

{GR-15310 Aghia Paraskevi, Attiki, Greece}

{$^\dagger$ Institute for Nuclear Research and Nuclear Energy, Bulgarian
Academy of Sciences }

{72 Tzarigrad Road, BG-1784 Sofia, Bulgaria}

\end{center}

\bigskip\bigskip
\centerline{\bf Abstract} \medskip
It is proved that the potentials of the form $\beta^{2n}$ (with $n$ being 
integer) provide a ``bridge'' between the U(5) symmetry of the 
Bohr Hamiltonian with a harmonic oscillator potential (occuring for $n=1$) 
and the E(5) model of Iachello (Bohr Hamiltonian with an infinite 
well potential, materialized for $n\to \infty$). Parameter-free (up to overall 
scale factors) predictions for spectra and B(E2) transition rates are 
given for the potentials $\beta^4$, $\beta^6$, $\beta^8$, corresponding to 
$R_4=E(4)/E(2)$ ratios of  2.093, 2.135, 2.157 respectively, compared 
to the $R_4$ ratios 2.000 of U(5) and 2.199 of E(5). Hints about nuclei 
showing this behaviour, as well as about potentials ``bridging'' the E(5) 
symmetry with O(6) are briefly discussed. A note about the appearance 
of Bessel functions in the framework of E(n) symmetries is given as a 
by-product. 
 

\newpage

{\bf 1. Introduction} 

The recently introduced E(5) \cite{IacE5} and X(5) \cite{IacX5} models 
are supposed to describe shape phase transitions in atomic nuclei, 
the former being related to the transition from U(5) (vibrational) 
to O(6) ($\gamma$-unstable) nuclei, and the latter corresponding to the 
transition from U(5) to SU(3) (prolate deformed) nuclei. In both cases 
the original Bohr collective Hamiltonian \cite{Bohr} is used, with an infinite 
well potential in the collective $\beta$-variable, after separating variables 
in two different ways. The selection of an infinite well potential in 
the $\beta$-variable in both cases is justified by the fact that the 
potential is expected to be flat around the point at which a shape phase 
transition occurs. In both models the predictions for nuclear spectra 
(normalized to the excitation energy of the first excited state) 
and B(E2) transition rates (normalized to the B(E2) transition rate 
connecting the first excited state to the ground state) do not contain 
any free parameters, thus providing useful benchmarks for nuclei 
in these two critical regions. 

In the present paper we study a sequence of potentials building a ``bridge''
between the U(5) symmetry of the Bohr Hamiltonian (corresponding to 
a five-dimensional (5-D) harmonic oscillator \cite{CM870}) 
and the E(5) model. 
The potentials, which are of the form $u_{2n}(\beta) = \beta^{2n}/2$, with 
$n$ being integer, are shown in Fig. 1.  
The Bohr Hamiltonian is obtained for $n=1$, while E(5) 
(which corresponds to an infinite well potential) 
occurs for $n\to \infty$ (in practice $n=4$ is already quite close to E(5)).  
Parameter-independent predictions for the spectra and B(E2) values 
(up to the overall scales mentioned above) are obtained for the potentials
$\beta^4$, $\beta^6$, $\beta^8$. 
In addition to providing a number of models giving predictions directly 
comparable to experiment, the present sequence of potentials shows the way 
for approaching the E(5) symmetry starting from U(5) and gives a hint 
on how to approach the E(5) symmetry starting from O(6). 

It should be pointed out that the $\beta^4$ potential has already 
received attention \cite{Arias}, since it turns out to be the critical 
potential for the U(5) to O(6) shape phase transition in the realm 
of an appropriate Interacting Boson Model \cite{IA} Hamiltonian. 

In Section 2 of the present paper a sequence of potentials providing 
a ``bridge'' between the U(5) model of Bohr \cite{Bohr,CM870} and the E(5) 
model of Iachello \cite{IacE5} is introduced. Numerical results for spectra
and B(E2) transition rates 
are shown in Sections 3 and 4 respectively, while Section 
5 contains a note on the appearance of Bessel functions in the framework 
of E(n). Perspectives for further experimental and theoretical work are 
discussed in Section 6, while in Section 7 the conclusions are summarized.  

{\bf 2. E(5), U(5), and a sequence of potentials between them} 
 
The original Bohr Hamiltonian \cite{Bohr} is
\begin{equation}\label{eq:e1}
H = -{\hbar^2 \over 2B} \left[ {1\over \beta^4} {\partial \over \partial 
\beta} \beta^4 {\partial \over \partial \beta} + {1\over \beta^2 \sin 
3\gamma} {\partial \over \partial \gamma} \sin 3 \gamma {\partial \over 
\partial \gamma} - {1\over 4 \beta^2} \sum_{k=1,2,3} {Q_k^2 \over \sin^2 
\left(\gamma - {2\over 3} \pi k\right) } \right] +V(\beta,\gamma),
\end{equation}
where $\beta$ and $\gamma$ are the usual collective coordinates describing the 
shape of the nuclear surface,
$Q_k$ ($k=1$, 2, 3) are the components of angular momentum, and $B$ is the 
mass parameter. 

Assuming that the potential depends only on the variable $\beta$, 
i.e. $V(\beta,\gamma) = U(\beta)$, one can proceed to separation of variables 
in the standard way \cite{Bohr, WJ1956}, using the wavefunction 
\begin{equation}\label{eq:e2} 
\Psi(\beta,\gamma, \theta_i) = f(\beta) \Phi(\gamma, \theta_i),
\end{equation}  
where $\theta_i$ $(i=1,2,3)$ are the Euler angles describing the orientation 
of the deformed nucleus in space. 

In the equation involving the angles, the eigenvalues of the second order 
Casimir operator of SO(5) occur, having the form 
 $\Lambda = \tau(\tau+3)$, where $\tau=0$, 1, 2, \dots is the quantum 
number characterizing the irreducible representations (irreps) of SO(5), 
called the ``seniority'' \cite{Rakavy}. This equation has been solved 
by Bes \cite{Bes}.   

The ``radial'' equation can be simplified by introducing \cite{IacE5} 
reduced energies $\epsilon = 
{2B\over \hbar^2} E$ and reduced potentials $u= {2B \over \hbar^2} U$, 
as well as by making the transformation  \cite{IacE5} 
$ \phi(\beta) = \beta^{3/2} f(\beta) $, leading to 
\begin{equation} \label{eq:e6} 
 \phi''+ {\phi'\over \beta} + \left[ \epsilon -u(\beta) -
{\left(\tau+{3\over 2}\right)^2 \over \beta^2} \right] \phi =0.
\end{equation} 

For $u(\beta)= \beta^2/2$ one obtains the original solution of Bohr 
\cite{Bohr,Dussel}, which corresponds to a 5-dimensional (5-D) harmonic 
oscillator 
characterized by the symmetry U(5) $\supset$ SO(5) $\supset$ SO(3) $\supset 
$SO(2)
\cite{CM870}, the eigenfunctions being proportional to 
Laguerre polynomials \cite{Mosh1555}
\begin{equation}\label{eq:e7}
F^\tau_\nu (\beta) = \left[ {2\nu! \over \Gamma \left(\nu+\tau +{5 \over 2}
\right)} \right]^{1\over 2} \beta^\tau  L_\nu^{\tau+{3\over 2}}(\beta^2)
e^{-\beta^2 /2},
\end{equation}
where $\Gamma(n)$ stands for the $\Gamma$-function, 
and the spectrum having the simple form  
\begin{equation}\label{eq:e8}
E_N = N+{5\over 2}, \qquad N=2\nu+\tau, \qquad \nu=0,1,2,3,\ldots 
\end{equation}

For $u(\beta)$ being a 5-D infinite well 
\begin{equation}\label{eq:e9}
  u(\beta) = \left\{ \begin{array}{ll} 0 & \mbox{if $\beta \leq \beta_W$} \\
\infty  & \mbox{for $\beta > \beta_W$} \end{array} \right.
\end{equation} 
one obtains the E(5) model of Iachello \cite{IacE5}, in which the 
eigenfunctions are 
Bessel functions $J_{\tau+3/2}(z)$ (with $z=\beta k$, $k =\sqrt{\epsilon}$), 
while the spectrum is determined by the zeros of the Bessel functions 
\begin{equation}\label{eq:e10}  
E_{\xi,\tau} = {\hbar^2 \over 2B} k^2_{\xi,\tau}, \qquad 
k_{\xi,\tau} = {x_{\xi,\tau} \over \beta_W}
\end{equation}
where $ x_{\xi,\tau}$ is the  $\xi$-th zero of the Bessel function 
$J_{\tau+3/2}(z)$. The relevant symmetry in this case is 
E(5)$\supset$SO(5)$\supset$SO(3)$\supset$SO(2), where the Euclidean algebra 
in 5 dimensions, E(5), is generated by the 5-D momenta $\pi_\mu$ and the 
5-D angular momenta $L_{\mu\nu}$, while SO(5) is generated by the 
$L_{\mu\nu}$ alone \cite{IacX5}. $\tau$, $L$, and $M$ are the 
quantum numbers characterizing the irreps of SO(5), SO(3), and SO(2)
respectively. The values of angular momentum $L$ contained in each 
irrep of SO(5) (i.e. for each value of $\tau$) are given by the 
algorithm \cite{IA} 
\begin{equation}\label{eq:e11} 
\tau = 3 \nu_\Delta +\lambda, \qquad \nu_\Delta =0, 1, \ldots, 
\end{equation}
\begin{equation}\label{eq:e12} 
L=\lambda, \lambda+1, \ldots, 2\lambda -2, 2\lambda 
\end{equation}
(with $2\lambda -1$ missing), 
where $\nu_\Delta$ is the missing quantum number in the reduction 
SO(5) $\supset$ SO(3), and are listed in Table 1. 

The spectra of the $u(\beta)=\beta^2/2$ potential and of the E(5) model 
become directly comparable by establishing the formal correspondence 
\begin{equation}\label{eq:e12a}
\nu= \xi-1.
\end{equation}
It should be emphasized that the quantum numbers appearing in Eq. 
(\ref{eq:e12a}) have different origins, $\nu$ being an oscillator 
quantum number labeling the number of zeros of a Laguerre polynomial,
while $\xi$ is labeling the order of a zero of a Bessel function. 
Eq. (\ref{eq:e12a}) establishes a formal one-to-one correspondence 
between the states in the two spectra and allows one to continue 
using for the states the notation $L_{\xi,\tau}$ (where $L$ is the 
angular momentum), as in Ref. \cite{IacE5}, although a notation 
$L_{\nu,\tau}$ would have been equally appropriate. The ground state 
band corresponds to $\xi=1$ (or, equivalently, $\nu=0$). 

The two cases mentioned above are the only ones in which Eq. (\ref{eq:e6})
is exactly soluble, giving spectra characterized by $R_4 = E(4)/E(2)$ 
ratios 2.00 and 2.20 respectively. However, the numerical solution of 
Eq. (\ref{eq:e6}) for potentials other than the ones mentioned above 
is a straightforward task \cite{Trolt}, in which one uses the chain 
U(5)$\supset$SO(5)$\supset$SO(3)$\supset$SO(2) for the classification 
of the states.  

Not all potentials can be used in Eq. (\ref{eq:e6}), though, since they have 
to obey the restrictions imposed by the 24 transformations mentioned 
in \cite{Bohr} and listed explicitly in \cite{Corr}. These restrictions allow 
the presence of even powers of $\beta$ in the potentials, while odd powers 
of $\beta$ should be accompanied by $\cos 3\gamma$ \cite{GM}. 

A particularly interesting sequence of potentials is given by 
\begin{equation}\label{eq:e13}
u_{2n}(\beta) = {\beta^{2n} \over 2},
\end{equation}
with $n$ being an integer. For $n=1$ the Bohr case (U(5)) is obtained, 
while for $n \to\infty$ the infinite well of E(5) is obtained
\cite{Bender}.  Therefore this sequence of potentials provides a ``bridge''
between the U(5) symmetry and the E(5) model, using their common 
SO(5)$\supset$SO(3) chain of subalgebras for the classification of the 
spectra.  

{\bf 3. Spectra} 

Numerical results for the spectra of the $\beta^4$, $\beta^6$, and $\beta^8$ 
potentials have been obtained through two different methods. In one 
approach, the representation of the position and momentum operators in matrix 
form \cite{Korsch} has been used, while in the other the direct integration 
method \cite{Scheid} has been applied. In the latter, the differential 
equation is solved for each value of $\tau=0,1,2,\ldots$ separately, the 
successive eigenvalues for each value of $\tau$ labeled by $\xi=1,2,3,\ldots$
(or, equivalently, by $\nu=0,1,2,\ldots$). The two methods give results 
mutually consistent, the second one appearing of more general applicability. 
The results are shown in Table 2, where excitation energies relative to the 
ground state, normalized to the excitation energy of the first excited state, 
are exhibited. 
 
In Table 2 the labels E(5)-$\beta^4$, E(5)-$\beta^6$, E(5)-$\beta^8$ have been 
used for the above-mentioned potentials, their meaning being that 
E(5)-$\beta^{2n}$ corresponds to the potential $\beta^{2n}/2$ plugged in 
the differential equation obtained in the framework of the E(5) model. 
In this notation E(5)-$\beta^2$ coincides with the original U(5) model 
of Bohr \cite{Bohr}, while E(5)-$\beta^{2n}$ with $n\to \infty$ is simply 
the original E(5) model \cite{IacE5}.

From Table 2 it is clear that in all bands and for all values of the 
angular momentum, $L$, the potentials $\beta^4$, $\beta^6$, $\beta^8$ 
gradually lead from the U(5) case to the E(5) results in a smooth way.    
The same conclusion is drawn from Fig. 2(a), where several levels of the 
ground state band of each model are shown vs. the angular momentum $L$, 
as well as from Fig. 2(b), where the bandheads of several excited bands 
are shown for each model as a function of the index $\xi$. 

It is instructive to compare the results obtained with the potentials 
of Eq. (\ref{eq:e13}) to the ones provided by the potentials \cite{IacE5,Roos}
\begin{equation}\label{eq:e14}
u(\beta)= {1\over 2} (1-\eta) \beta^2 + {\eta\over 4} (1-\beta^2)^2,
\end{equation}
where $\eta$ is a control parameter. Results for the spectra of these 
potentials (for $\eta= 1/4$, 1/2, 3/4, 1) are shown in Table 3, while 
for $\eta=0$ it is clear that the Bohr U(5) case is reproduced. The following 
observations can be made:

1) For $\eta=1/2$ the results coincide with these of E(5)-$\beta^4$, as 
expected, since for $\eta =1/2$ Eq. (\ref{eq:e14}) gives $u(\beta)
= (\beta^4+1)/8$, while in Tables 2 and 3 excitation energies relative to the 
ground state and normalized to the excitation energy of the first excited 
state are shown. 
 
2) Giving to the control parameter $\eta$ the values 0, 1/4, 1/2, 3/4, 1, 
one obtains spectra characterized by $R_4$ ratios 2.00, 2.06, 2.09, 
2.11, 2.13 respectively. 

3) It has been noticed in \cite{IacE5} that the potential should exhibit 
a flat behaviour when the system undergoes a phase transition. 
The only flat potential contained in the family of potentials of Eq. 
(\ref{eq:e14}) is the above mentioned potential $u(\beta)=(b^4+1)/8$, 
which occurs for $\eta=1/2$, giving $R_4=2.09$~. In contrast, the sequence 
of potentials given in Eq. (\ref{eq:e13}) is indeed a series of gradually 
flatter (with increasing $n$) potentials, giving the infinite well potential
of E(5) (with $R_4=2.20$) as a limiting case. These potentials therefore 
do provide a complete ``bridge'' between U(5) and E(5).

{\bf 4. B(E2) transition rates} 

In nuclear structure it is well known that electromagnetic transition rates 
are quantities sensitive to the details of the underlying microscopic 
structure, as well as to details of the theoretical models, much more 
than the corresponding spectra. It is therefore a must to calculate B(E2) 
ratios (normalized to B(E2:$2_1^+\to 0_1^+$)=100) for the potentials 
of Eq. (\ref{eq:e13}). 

The quadrupole operator has the form \cite{WJ1956} 
\begin{equation}\label{eq:e15}
T^{(E2)}_\mu= t \alpha_\mu = 
t \beta \left[ {\cal D}^{(2)}_{\mu,0}(\theta_i) \cos\gamma  +{1\over 
\sqrt{2}} ({\cal D} ^{(2)}_{\mu,2} (\theta_i)+ {\cal D}^{(2)}_{\mu,-2} 
(\theta_i) ) \sin\gamma
\right],
\end{equation}
where $t$ is a scale factor and ${\cal D}(\theta_i)$ denote Wigner functions 
of the Euler angles, while the B(E2) transition rates are given by 
$$
B(E2; \varrho_i L_i\to \varrho_f L_f) =
\frac{1}{2L_i+1}\,|\langle \varrho_f L_f||T^{(E2)}||\varrho_i L_i\rangle|^2 
$$
\begin{equation}\label{eq:e16}
= \frac{2L_f+1}{2L_i+1}\, B(E2; \varrho_f L_f\to \varrho_i L_i), 
\end{equation}
where by $\varrho$ quantum numbers other than the angular momentum $L$ are 
denoted. 

The states with $\nu_\Delta =0$ and $L=2\tau$ can be written in the form
dictated by Eq. (\ref{eq:e2}) 
\begin{equation}\label{eq:e31}
|\xi, \tau, \nu_\Delta=0, L=2\tau, M=L\rangle =  f_{\xi\tau}(\beta)\; 
\Phi^{\tau,\; \nu_\Delta=0}_{L=2\tau,\; M=L}(\gamma,\theta_i) =
f_{\xi\tau}(\beta)\; \phi_{\tau}(\gamma,\theta_i), 
\end{equation}
where the functions $\phi_{\tau}(\gamma,\theta_i)$ have the form \cite{Bes} 
\begin{equation}\label{eq:e32}
\phi_{\tau}(\gamma,\theta_i) = \frac{1}{\sqrt{A_{\tau}}}\,
\left(\frac{\alpha_2}{\beta}\right)^\tau,
\end{equation}
with $\alpha_2$ defined in Eq. (\ref{eq:e15}) and with the normalization factor
\begin{equation}\label{eq:e33}
A_{\tau} = \frac{\tau!}{(2\tau+3)!!}\,(4\pi)^2
\end{equation}
determined from the normalization condition
\begin{equation}\label{eq:e34} 
\int\limits_{\gamma=0}^{\frac{\pi}{3}}\!\!\int
\phi^\ast_{\tau}(\gamma,\theta_i)\,\phi_{\tau}(\gamma,\theta_i)
\,\sin3\gamma\,\sin\theta_2\,d\gamma\, d\theta_1\, d\theta_2\, d\theta_3 = 1.
\end{equation}

From Eqs. (\ref{eq:e16}) and (\ref{eq:e32}) one obtains
\begin{equation}\label{eq:e35}
B(E2; L_{\xi,\tau}\to (L+2)_{\xi',\tau+1}) =
\frac{(\tau+1)(4\tau+5)}{(2\tau+5)(4\tau+1)}\;t^2\,I^2_{\xi',\tau+1;\;\xi,\tau}
\qquad;\quad L=2\tau, 
\end{equation}
\begin{equation}\label{eq:e36} 
B(E2; (L+2)_{\xi',\tau+1}\to L_{\xi,\tau}) = 
\frac{\tau+1}{2\tau+5}\;t^2\,I^2_{\xi',\tau+1;\;\xi,\tau}
\qquad\qquad\qquad;\quad L=2\tau, 
\end{equation}
where
\begin{equation}\label{eq:e37}
I_{\xi',\tau+1;\;\xi,\tau} = \int_{0}^{\infty}\beta f_{\xi'\,\tau+1}
(\beta)\,f_{\xi\tau}(\beta)\beta^4 d\beta. 
\end{equation}

In the special case of the potential being a 5-D infinite well the 
eigenfunctions are
\begin{equation}\label{eq:e38}
f_{\xi\tau}(\beta) = \frac{1}{\sqrt{C_{\xi,\tau}}}\,
\beta^{-3/2} J_{\tau+3/2}\Bigl(x_{\xi,\tau}\,\frac{\beta}{\beta_W}\Bigr),
\end{equation} 
with
\begin{equation}\label{eq:e39} 
C_{\xi,\tau} = \frac{\beta_W^2}{2}J_{\tau+5/2}^2(x_{\xi,\tau}), 
\end{equation}
where $x_{\xi,\tau}$ is the  $\xi$-th zero of the Bessel function 
$J_{\tau+3/2}(z)$, while the constants $C_{\xi,\tau}$ are obtained from the 
normalization condition
\begin{equation}\label{eq:e40} 
\int_{0}^{\beta_W} f^2_{\xi \tau}(\beta)\; \beta^4 d\beta = 1. 
\end{equation}
In this case the integrals of Eq. (\ref{eq:e37}) take the form
$$
I_{\xi',\tau+1;\;\xi,\tau} = \int_{0}^{\beta_W}\beta f_{\xi'\,\tau+1}
(\beta)\,f_{\xi\tau}(\beta)\beta^4 d\beta
$$
\begin{equation}\label{eq:e41}
= (C_{\xi',\tau+1} C_{\xi,\tau})^{-1/2} \beta_W^3
\int_{0}^{1} z^2 J_{\tau+5/2}(x_{\xi',\tau+1}\, z)\,
J_{\tau+3/2}(x_{\xi,\tau}\, z)\,dz. 
\end{equation}

In the case of the oscillator potential $u(\beta)=\beta^2/2$ the eigenfunctions
are given by Eq. (\ref{eq:e7}). In this case the integrals 
$I_{n',\tau+1;\;n,\tau}$ appearing in Eq.  (\ref{eq:e37})
(where $n=\xi-1$ and $n'=\xi'-1$, as mentioned in Eq. \ref{eq:e12a})~)
in the cases $n'=n, n\pm1$ are found to be
\begin{equation}\label{eq:e42}
I_{n,\tau+1;\; n,\tau} = \sqrt{n+\tau+5/2},
\end{equation}
\begin{equation}\label{eq:e43}
I_{n+1,\tau+1;\; n,\tau} = 0, \qquad\qquad\qquad n\geq 0, 
\end{equation}
\begin{equation}\label{eq:e44} 
I_{n-1,\tau+1;\; n,\tau} = \sqrt{n}, \qquad\qquad n\geq 1,
\end{equation}
leading to the ratios  
\begin{equation}\label{eq:e45} 
\frac{B(E2; (L+2)_{\xi,\tau+1}\to L_{\xi,\tau})}{B(E2; 2_{1,1}\to 0_{1,0})} =
\frac{(\tau+1)}{(2\tau+5)}\;(2\xi + 2\tau +3),
\quad L=2\tau, \ \xi\geq1,
\end{equation}
\begin{equation}\label{eq:e46}
\frac{B(E2; L_{\xi,\tau}\to (L+2)_{\xi-1,\tau+1})}{B(E2; 2_{1,1}\to 0_{1,0})} 
=\frac{(\tau+1)(4\tau+5)}{(2\tau+5)(4\tau+1)}\;(2\xi-2),
\quad L=2\tau, \ \xi\geq2.
\end{equation}

The results of the calculations for intraband transitions are shown 
in Table 4, while interband transitions are shown in Table 5. In addition, 
the normalized B(E2) transition rates within the ground state band 
of each model are shown in Fig. 2(c). In all cases 
a smooth evolution from U(5) to E(5) is seen. Furthermore, the results 
are in agreement to general qualitative expectations: the more rotational  
the nucleus, the less rapid the increase (with increasing initial angular 
momentum) of the B(E2)s within the ground state band should be
(in the absence of bandcrossings). 
Indeed the most rapid increase is seen in the case of U(5), while the 
slowest increase is observed in the case of E(5). 
The E(5) results reported in Tables 4 and 5 are in good agreement 
with the results given in Ref. \cite{IacTri}. 

Finally, in Fig. 3 the lowest part of the spectrum, which is of interest 
for comparisons with experimental data, together with all 
relevant B(E2) transition rates is shown for the models E(5)-$\beta^4$, 
E(5)-$\beta^6$ and E(5)-$\beta^8$. The models E(5) and U(5) are also included 
for comparison. 

{\bf 5. A note on E(n) and Bessel functions} 

Concerning the appearance of Bessel functions in the case of E(5), the 
following mathematical remarks can be made in the general case of the 
Euclidean 
algebra in $n$ dimensions, E(n), which is the semidirect sum \cite{Wyb} 
of the algebra T$_n$ of translations in $n$ dimensions, generated 
by the momenta
\begin{equation}\label{eq:e17} 
P_j = -i {\partial \over \partial x_j}, 
\end{equation} 
 and the SO(n) algebra 
of rotations in $n$ dimensions, generated by the angular momenta
\begin{equation}\label{eq:e18} 
 L_{jk} =-i \left(x_j{\partial \over \partial x_k} -x_k {\partial \over
\partial x_j} \right), 
\end{equation}
symbolically written as E(n) = T$_{\rm n}$ $\oplus_s$ SO(n) \cite{Barut}.  
The generators of E(n) satisfy the commutation relations 
\begin{equation}\label{eq:e19} 
 [P_i, P_j] =0, \qquad [P_i, L_{jk}] = i ( \delta_{ik} P_j - \delta_{ij} P_k),
\end{equation}
\begin{equation}\label{eq:e20} 
 [L_{ij}, L_{kl}]=i (\delta_{ik} L_{jl} +\delta_{jl} L_{ik} 
-\delta_{il} L_{jk} -\delta_{jk} L_{il}).
\end{equation}
From these commutation relations one can see that the square of the total 
momentum, $P^2$, is a second order Casimir operator of the algebra, while 
the eigenfunctions of this operator satisfy the equation 
\begin{equation}\label{eq:e21} 
 \left( -{1\over r^{n-1}} {\partial \over \partial r} r^{n-1} {\partial \over 
\partial r} + { \omega(\omega+n-2) \over r^2} \right) F(r) = k^2 F(r), 
\end{equation} 
in the left hand side of which the eigenvalues of the Casimir operator 
of SO(n), $\omega(\omega+n-2)$ appear \cite{Mosh1555}. 
Putting 
\begin{equation}\label{eq:e22} 
 F(r) = r^{(2-n)/2} f(r),
\end{equation}
and 
\begin{equation}\label{eq:e23} 
 \nu= \omega+{n-2\over 2},
\end{equation}
Eq. (\ref{eq:e21}) is brought into the form 
\begin{equation}\label{eq:e24} 
 \left( {\partial^2 \over \partial r^2} + {1\over r} {\partial \over \partial 
r} + k^2 - { \nu^2 \over r^2}\right) f(r)  =0, 
\end{equation}
the eigenfunctions of which are the Bessel functions $f(r) = J_\nu(kr) $
\cite{AbrSt}. We see therefore that the Bessel functions appear in general 
in this type of problems 
when the potential is vanishing, so that only the kinetic energy term appears 
in the Hamiltonian.  

A similar result for the case of the $n$-dimensional harmonic oscillator 
has been obtained in Ref. \cite{Rakavy} and developed in more detail 
in Ref. \cite{Mosh1555}, showing the appearance of Laguerre polynomials 
in the eigenfunctions of the harmonic oscillator in all dimensions.  

{\bf 6. Perspectives} 

It is interesting to examine if there is any experimental evidence supporting 
the E(5)-$\beta^{2n}$ predictions. It is clear that the first regions 
to be considered are the ones around the nuclei which have been identified 
as good candidates for E(5), i.e. $^{134}$Ba \cite{Ba134}, $^{104}$Ru 
\cite{Ru104}, $^{102}$Pd \cite{Pd102}. A very preliminary search indicates 
that $^{98}$Ru \cite{Ru98} can be a candidate for E(5)-$\beta^6$, 
while $^{100}$Pd \cite{Pd100} can be a candidate for E(5)-$\beta^4$. 
Existing data for the ground state bands of these 
nuclei are compared to the theoretical predictions in Table 6. However, 
much more detailed information on the spectra and B(E2) transitions of these 
nuclei are required before final conclusions can be reached.   

Concerning future theoretical work, at least two directions open up: 

1) One should study a similar sequence of potentials serving as a ``bridge'' 
between U(5) and X(5) \cite{IacX5}. This task has been carried out in 
Ref. \cite{X5}. 

2) One should try to find a sequence of potentials interpolating 
between O(6) and E(5), as well as between SU(3) and X(5). In other words, 
one should try to approach E(5) and X(5) ``from the other side''. 
From the classical limit of the O(6) and SU(3) symmetries 
of the Interacting 
Boson Model \cite{IA} it is clear that for this purpose potentials with a 
minimum at $\beta \neq 0$ should be considered, the Davidson-like  potentials 
\cite{Dav} 
\begin{equation}\label{eq:e25}  
u_{2n}^D(\beta) = \beta^{2n} + {\beta_0^{4n} \over \beta^{2n} }
\end{equation}
being strong candidates. The Davidson potential, corresponding to $n=1$, 
is known to be exactly soluble \cite{Dav,Rowe}. Work in these directions 
is in progress. 

3) Another candidate for the task described in 2) is the sextic oscillator
with a centrifugal barrier \cite{Ushv}, recently considered in Ref. 
\cite{Levai}, which contains $\beta^2$, $\beta^4$, $\beta^6$ and 
$\beta^{-2}$ terms with coefficients interrelated in an appropriate 
way in order to quarantee that the potential is a Quasi-Exactly 
Soluble one \cite{TU,TurbCMP,TurbSP}. The sextic oscillator with a 
centrifugal barrier contains two free parameters, and it is 
capable of producing potentials with minima at $\beta\neq 0$ \cite{Levai}. 

{\bf 7. Conclusion} 

It has been proved that the potentials $\beta^{2n}$ (with $n$ being integer) 
provide a complete ``bridge'' between the U(5) symmetry of the Bohr 
Hamiltonian with a harmonic oscillator potential (occuring for $n=1$) 
and the E(5) model of Iachello, which is obtained from the Bohr Hamiltonian 
when an infinite well potential is plugged in it (materialized for 
$n \to \infty$). 
Parameter-free (up to overall scale factors) predictions for spectra 
and B(E2) transition rates have been given for the potentials $\beta^4$, 
$\beta^6$, $\beta^8$, called the E(5)-$\beta^4$, E(5)-$\beta^6$, and 
E(5)-$\beta^8$ models, respectively. Hints about nuclei showing this behaviour,
as well as about potentials approaching E(5) ``from the other side''
(i.e. providing a ``bridge'' between O(6) and E(5)) have been briefly 
discussed. A mathematical note on the appearance of Bessel functions 
in the framework of E(n) models has been given as a by-product. 

{\bf Acknowledgements} 

The authors are thankful to Rick Casten (Yale), Jean Libert (Orsay), and 
Werner Scheid (Giessen) for illuminating discussions. Support through the 
NATO Collaborative Linkage Grant PST.CLG 978799 is gratefully acknowledged.

\bigskip

\centerline{\bf Figure captions} 

\bigskip
{\bf Fig. 1.} The potentials $\beta^{2n}$, with $n=1$ (harmonic oscillator, 
solid line), $n=2$ (dash line), $n=3$ (dash dot), $n=4$ (dot), $n=8$ 
(das dot dot), $n=16$ (short dash dot), $n=32$ (short dot), gradually 
approaching (with increasing $n$)  the infinite well potential.  

\medskip
{\bf Fig. 2} (Color online) (a) Levels of the ground state bands of the models 
E(5)-$\beta^{2n}$ with $n=2$-4 and of the U(5) and E(5) models, vs. the 
angular 
momentum $L$. In each model all levels are normalized to the energy 
of the first excited state. See Section 3 for further discussion. 
(b) Bandhead energies of excited bands of the same models and with the 
same normalization, vs. the  band index $\xi$. See Section 3 for 
further discussion. (c) B(E2:$L_f+2 \to L_f$) transition rates within the 
ground state bands of the same models, vs. the angular momentum of the 
final state, $L_f$. In each model all rates are normalized to the one 
between the lowest states, B(E2:$2\to 0$). See Section 4 for further 
discussion. 

\medskip
{\bf Fig. 3} Lowest part of the spectrum, together with the relevant B(E2) 
transition rates, for the models U(5) (labeled as $\beta^2$), 
E(5)-$\beta^4$ (labeled as $\beta^4$), E(5)-$\beta^6$ ($\beta^6$), 
E(5)-$\beta^8$ ($\beta^8$), and E(5) (labeled as infinite well). 
See Section 4 for further discussion. The results for E(5)-$\beta^4$ 
and E(5) compare well with prior work reported in Refs. \cite{Arias} 
and \cite{IacTri} respectively.   

\newpage 

\begin{table}

\caption{Quantum numbers appearing in the SO(5)$\supset$SO(3) reduction 
\cite{IA}, occuring from Eqs. (\ref{eq:e11}) and (\ref{eq:e12}).  
}

\bigskip

\begin{tabular}{l l l l}
\hline
$\tau$ & $\nu_\Delta$ & $\lambda$ & $L$         \\
       &              &           &             \\  
0      &  0           &     0     &  0          \\
       &              &           &             \\
1      &  0           &     1     &  2          \\
       &              &           &             \\
2      &  0           &     2     &  4,2        \\
       &              &           &             \\
3      &  0           &     3     &  6,4,3      \\
3      &  1           &     0     &  0          \\
       &              &           &             \\
4      &  0           &     4     &  8,6,5,4    \\
4      &  1           &     1     &  2          \\
       &              &           &             \\  
5      &  0           &     5     & 10,8,7,6,5  \\
5      &  1           &     2     & 4,2         \\
       &              &           &             \\
6      &  0           &     6     & 12,10,9,8,7,6\\
6      &  1           &     3     & 6,4,3       \\
6      &  2           &     0     & 0           \\
\hline 
\end{tabular}
\end{table}

\newpage 

\begin{table}

\caption{Spectra of the E(5)-$\beta^4$, E(5)-$\beta^6$, and E(5)-$\beta^8$ 
models, compared to the predictions of the U(5) (Eq. (\ref{eq:e8})) 
and E(5) (Eq. (\ref{eq:e10})) models. For each value of $\tau$, only 
the maximum value of $L$ occuring for it, $L_{max}$, is reported.
The rest of the allowed values of $L$ for each value of $\tau$, indicating 
states having the same energy as the state with $L_{max}$, can be read from 
Table 1. The lowest four levels in each band of E(5)-$\beta^4$ are in good 
agreement with results already published in Ref. \cite{Arias}. 
}

\bigskip

\begin{tabular}{r r r r r r r r}
\hline
band & $\tau$ & $L_{max}$ & U(5) & E(5)-$\beta^4$ & E(5)-$\beta^6$ 
& E(5)-$\beta^8$ & E(5) \\
\hline
$\xi=1$ & & &       &       &       &       &       \\
 & 0& 0  & 0.000 & 0.000 & 0.000 & 0.000 & 0.000 \\
 & 1& 2  & 1.000 & 1.000 & 1.000 & 1.000 & 1.000 \\
 & 2& 4  & 2.000 & 2.093 & 2.135 & 2.157 & 2.199 \\
 & 3& 6  & 3.000 & 3.265 & 3.391 & 3.459 & 3.590 \\
 & 4& 8  & 4.000 & 4.508 & 4.757 & 4.894 & 5.169 \\
 & 5& 10 & 5.000 & 5.813 & 6.225 & 6.456 & 6.934 \\
 & 6& 12 & 6.000 & 7.176 & 7.788 & 8.138 & 8.881 \\
 & 7& 14 & 7.000 & 8.592 & 9.442 & 9.935 &11.009 \\
 & 8& 16 & 8.000 &10.057 &11.180 &11.841 &13.316 \\
 & 9& 18 & 9.000 &11.569 &13.000 &13.854 &15.799 \\ 
 &10& 20 &10.000 &13.124 &14.898 &15.968 &18.459 \\
 &11& 22 &11.000 &14.720 &16.871 &18.182 &21.294 \\
 &12& 24 &12.000 &16.355 &18.916 &20.492 &24.302 \\
 &13& 26 &13.000 &18.028 &21.031 &22.896 &27.484 \\
 &14& 28 &14.000 &19.737 &23.213 &25.391 &30.837 \\ 
 &15& 30 &15.000 &21.480 &25.460 &27.975 &34.363 \\
\hline
$\xi=2$ & & &        &          &          &          &       \\
 & 0& 0 & 2.000 & 2.390 & 2.619 & 2.756 & 3.031\\
 & 1& 2 & 3.000 & 3.625 & 4.012 & 4.255 & 4.800\\
 & 2& 4 & 4.000 & 4.918 & 5.499 & 5.874 & 6.780\\
 & 3& 6 & 5.000 & 6.266 & 7.075 & 7.607 & 8.967\\
 & 4& 8 & 6.000 & 7.666 & 8.738 & 9.450 &11.357\\
 & 5&10 & 7.000 & 9.115 &10.483 &11.400 &13.945\\
\hline
$\xi=3$ & & &          &         &         &           &        \\
 & 0& 0 & 4.000 & 5.153 & 5.887 & 6.364 & 7.577\\
 & 1& 2 & 5.000 & 6.563 & 7.588 & 8.269 &10.107\\
 & 2& 4 & 6.000 & 8.015 & 9.363 &10.274 &12.854\\
 & 3& 6 & 7.000 & 9.509 &11.213 &12.379 &15.814\\
 & 4& 8 & 8.000 &11.043 &13.134 &14.580 &18.983\\  
 & 5&10 & 9.000 &12.617 &15.125 &16.875 &22.359\\
\hline
$\xi=4$ & & &         &           &         &            &        \\
 & 0& 0 & 6.000 & 8.213 & 9.698 &10.707 &13.639\\  
 & 1& 2 & 7.000 & 9.764 &11.661 &12.966 &16.928\\
 & 2& 4 & 8.000 &11.349 &13.687 &15.316 &20.436\\
 & 3& 6 & 9.000 &12.967 &15.776 &17.753 &24.161\\
 & 4& 8 &10.000 &14.619 &17.928 &20.278 &28.100\\
 & 5&10 &11.000 &16.304 &20.141 &22.888 &32.250\\
\hline 
\end{tabular}
\end{table}

\newpage

\begin{table}

\caption{Same as Table 2, but for spectra of the potentials of Eq. 
(\ref{eq:e14}) for different values 
of the control parameter $\eta$, compared to the predictions of the U(5) 
($\eta =0$, Eq. (\ref{eq:e8})) and E(5) (Eq. (\ref{eq:e10}) models. 
}

\bigskip

\begin{tabular}{r r r r r r r r r} 
\hline
  &  & $\eta$ & 0 & 1/4   & 1/2   & 3/4   & 1     &  E(5) \\
band & $\tau$ & $L_{max}$ &   &       &       &       &       &       \\
\hline
$\xi=1$ & & &       &       &       &       &       &       \\
 & 0& 0  & 0.000 & 0.000 & 0.000 & 0.000 & 0.000 & 0.000 \\
 & 1& 2  & 1.000 & 1.000 & 1.000 & 1.000 & 1.000 & 1.000 \\
 & 2& 4  & 2.000 & 2.063 & 2.093 & 2.114 & 2.130 & 2.199 \\
 & 3& 6  & 3.000 & 3.183 & 3.265 & 3.323 & 3.368 & 3.590 \\
 & 4& 8  & 4.000 & 4.353 & 4.508 & 4.615 & 4.700 & 5.169 \\
 & 5& 10 & 5.000 & 5.569 & 5.813 & 5.982 & 6.115 & 6.934 \\
 & 6& 12 & 6.000 & 6.828 & 7.176 & 7.415 & 7.605 & 8.881 \\
 & 7& 14 & 7.000 & 8.127 & 8.592 & 8.911 & 9.164 &11.009 \\
 & 8& 16 & 8.000 & 9.462 & 10.057 &10.464 &10.786 &13.316 \\
 & 9& 18 & 9.000 &10.833 & 11.569 &12.071 &12.467 &15.799 \\ 
 &10& 20 &10.000 &12.238 & 13.124 &13.727 &14.204 &18.459 \\
 &11& 22 &11.000 &13.674 & 14.720 &15.432 &15.994 &21.294 \\
 &12& 24 &12.000 &15.140 & 16.355 &17.181 &17.833 &24.302 \\
 &13& 26 &13.000 &16.636 & 18.028 &18.973 &19.719 &27.484 \\
 &14& 28 &14.000 &18.159 & 19.737 &20.807 &21.651 &30.837 \\ 
 &15& 30 &15.000 &19.709 & 21.480 &22.679 &23.625 &34.363 \\
\hline
$\xi=2$ & & &        &          &          &          &       \\
 & 0& 0 & 2.000 & 2.251 & 2.390 & 2.498 & 2.590 & 3.031\\
 & 1& 2 & 3.000 & 3.419 & 3.625 & 3.776 & 3.902 & 4.800\\
 & 2& 4 & 4.000 & 4.629 & 4.918 & 5.124 & 5.292 & 6.780\\
 & 3& 6 & 5.000 & 5.881 & 6.266 & 6.537 & 6.756 & 8.967\\
 & 4& 8 & 6.000 & 7.171 & 7.666 & 8.011 & 8.287 &11.357\\
 & 5&10 & 7.000 & 8.497 & 9.115 & 9.542 & 9.883 &13.945\\
\hline
$\xi=3$ & & &          &         &         &           &        \\
 & 0& 0 & 4.000 & 4.785 & 5.153 & 5.419 & 5.639 & 7.577\\
 & 1& 2 & 5.000 & 6.082 & 6.563 & 6.905 & 7.182 &10.107\\
 & 2& 4 & 6.000 & 7.412 & 8.015 & 8.438 & 8.779 &12.854\\
 & 3& 6 & 7.000 & 8.774 & 9.509 &10.020 &10.430 &15.814\\
 & 4& 8 & 8.000 &10.166 &11.043 &11.649 &11.133 &18.983\\  
 & 5&10 & 9.000 &11.589 &12.617 &13.324 &13.887 &22.359\\
\hline
$\xi=4$ & & &         &           &         &            &        \\
 & 0& 0 & 6.000 & 7.548 & 8.213 & 8.681 & 9.061 &13.639\\  
 & 1& 2 & 7.000 & 8.951 & 9.764 &10.331 &10.788 &16.928\\
 & 2& 4 & 8.000 &10.382 &11.349 &12.019 &12.556 &20.436\\
 & 3& 6 & 9.000 &11.839 &12.967 &13.745 &14.366 &24.161\\
 & 4& 8 &10.000 &13.322 &14.619 &15.509 &16.218 &28.100\\
 & 5&10 &11.000 &14.831 &16.304 &17.311 &18.111 &32.250\\
\hline 
\end{tabular}
\end{table}

\newpage

\begin{table}

\caption{Intraband B(E2) transition rates for the E(5)-$\beta^4$,
E(5)-$\beta^6$, and E(5)-$\beta^8$ models, compared to the predictions
of the U(5) and E(5) models. Some of the E(5)-$\beta^4$ transitions 
have been reported in Ref. \cite{Arias}, with very similar values. 
See Section 4 for details. 
}

\bigskip

\begin{tabular}{l r r r r r r r}
\hline
bands & $(L_{\xi,\tau})_i$ & $(L_{\xi,\tau})_f$ & U(5) & E(5)-$\beta^4$
& E(5)-$\beta^6$ & E(5)-$\beta^8$ & E(5) \\
\hline
$(\xi=1)\to (\xi=1)$ &   &      &       &       &       &       &       \\
    & $2_{1,1}$& $0_{1,0}$& 100.00 & 100.00 & 100.00 & 100.00  & 100.00 \\
    & $4_{1,2}$& $2_{1,1}$& 200.00 & 183.20 & 176.60 & 173.32  & 167.40 \\
    & $6_{1,3}$& $4_{1,2}$& 300.00 & 256.37 & 239.80 & 231.64  & 216.88\\
    & $8_{1,4}$& $6_{1,3}$& 400.00 & 322.73 & 294.27 & 280.39  & 255.20\\
    &$10_{1,5}$& $8_{1,4}$& 500.00 & 384.12 & 342.57 & 322.51  & 286.01\\
& $12_{1,6} $ & $10_{1,5}$ & 600.00 & 441.65 & 386.26 & 359.74 & 311.47 \\
& $14_{1,7} $ & $12_{1,6}$ & 700.00 & 496.11 & 426.36 & 393.25 & 332.95 \\
& $16_{1,8} $ & $14_{1,7}$ & 800.00 & 548.02 & 463.57 & 423.80 & 351.39 \\
& $18_{1,9} $ & $16_{1,8}$ & 900.00 & 597.78 & 498.40 & 451.94 & 367.44 \\
& $20_{1,10}$ & $18_{1,9}$ &1000.00 & 645.69 & 531.23 & 478.10 & 381.56 \\
& $22_{1,11}$ & $20_{1,10}$ &1100.00 & 692.00 & 562.35 & 502.58 & 394.10 \\
& $24_{1,12}$ & $22_{1,11}$ &1200.00 & 736.89 & 592.00 & 525.62 & 405.34 \\
& $26_{1,13}$ & $24_{1,12}$ &1300.00 & 780.52 & 620.35 & 547.41 & 415.48 \\
& $28_{1,14}$ & $26_{1,13}$ &1400.00 & 823.01 & 647.55 & 568.12 & 424.68 \\
& $30_{1,15}$ & $28_{1,14}$ &1500.00 & 864.47 & 673.73 & 587.86 & 433.09 \\
& $2_{1,2}$   & $2_{1,1}$   & 200.00 & 183.20 & 176.60 & 173.32 & 167.40 \\
& $4_{1,3}$   & $2_{1,2}$   & 157.14 & 134.29 & 125.61 & 121.34 & 113.60 \\
& $4_{1,3}$   & $4_{1,2}$   & 142.86 & 122.08 & 114.19 & 110.31 & 103.28 \\
& $3_{1,3}$   & $2_{1,2}$   & 214.29 & 183.12 & 171.29 & 165.46 & 154.91 \\
& $3_{1,3}$   & $4_{1,2}$   &  85.71 &  73.25 &  68.51 &  66.18 &  61.97 \\
& $0_{1,3}$   & $2_{1,2}$   & 300.00 & 256.37 & 239.80 & 231.64 & 216.88 \\
\hline
$(\xi=2) \to (\xi=2)$ &   &      &       &       &       &       &       \\
    & $2_{2,1}$& $0_{2,0}$& 140.00 & 112.64 &  98.97 &  91.24  &  75.22 \\
    & $4_{2,2}$& $2_{2,1}$& 257.14 & 197.92 & 170.97 & 156.06  & 124.32 \\
    & $6_{2,3}$& $4_{2,2}$& 366.67 & 271.04 & 230.57 & 208.71  & 161.52 \\
    & $8_{2,4}$& $6_{2,3}$& 472.73 & 336.84 & 282.53 & 253.85  & 191.58 \\
    &$10_{2,5}$& $8_{2,4}$& 576.92 & 397.56 & 329.12 & 293.70  & 216.77 \\
    & $2_{2,2}$& $2_{2,1}$& 257.14 & 197.92 & 170.97 & 156.06  & 124.32 \\
\hline
$(\xi=3) \to (\xi=3)$ &   &      &       &       &       &       &       \\
& $ 2_{3,1}$ & $0_{3,0}$ & 180.00 & 126.58 & 103.69 &  91.64 &  65.73 \\
& $ 4_{3,2}$ & $2_{3,1}$ & 314.29 & 214.91 & 173.97 & 152.67 & 106.63 \\
& $ 6_{3,3}$ & $4_{3,2}$ & 433.33 & 288.38 & 230.96 & 201.40 & 137.44 \\
& $ 8_{3,4}$ & $6_{3,3}$ & 545.45 & 353.71 & 280.48 & 243.22 & 162.57 \\
& $10_{3,5}$ & $8_{3,4}$ & 653.85 & 413.72 & 325.01 & 280.39 & 183.95 \\
& $2_{3,2}$  & $2_{3,1}$ & 314.29 & 214.91 & 173.97 & 152.67 & 106.63 \\
\hline
$(\xi=4) \to (\xi=4)$ &   &      &       &       &       &       &       \\
& $ 2_{4,1}$ & $0_{4,0}$ & 220.00 & 140.44 & 109.56 &  94.03 &  60.68 \\
& $ 4_{4,2}$ & $2_{4,1}$ & 371.43 & 232.42 & 179.66 & 153.33 &  96.89 \\
& $ 6_{4,3}$ & $4_{4,2}$ & 500.00 & 306.70 & 235.08 & 199.63 & 123.79 \\
& $ 8_{4,4}$ & $6_{4,3}$ & 618.18 & 371.85 & 282.79 & 239.04 & 145.70 \\
& $10_{4,5}$ & $8_{4,4}$ & 730.77 & 431.31 & 325.60 & 274.06 & 164.42 \\
\hline
\end{tabular}
\end{table}

\newpage

\begin{table}

\caption{Same as Table 4, but for interband B(E2) transitions.
}

\bigskip

\begin{tabular}{l r r r r r r r}
\hline
bands & $(L_{\xi,\tau})_i$ & $(L_{\xi,\tau})_f$ & U(5) & E(5)-$\beta^4$
& E(5)-$\beta^6$ & E(5)-$\beta^8$ & E(5) \\
\hline
$(\xi=2) \to (\xi=1)$ &   &      &       &       &       &       &       \\
& $0_{2,0}$ & $2_{1,1}$ & 200.00 & 141.77  & 118.98  & 107.57  &  86.79  \\
& $2_{2,1}$ & $4_{1,2}$ & 102.86 &  66.10  &  52.62  &  46.00  &  33.82  \\
& $2_{2,1}$ & $0_{1,0}$ &   0.00 &   0.16  &   0.30  &   0.38  &   0.47  \\
& $4_{2,2}$ & $6_{1,3}$ &  96.30 &  57.33  &  43.78  & 37.263  &  25.17  \\
& $4_{2,2}$ & $2_{1,1}$ &   0.00 &   0.24  &   0.45  &   0.56  &   0.69  \\
& $6_{2,3}$ & $8_{1,4}$ &  95.11 &  53.20  &  39.26  &  32.68  &  20.44  \\
& $6_{2,3}$ & $4_{1,2}$ &   0.00 &   0.28  &   0.52  &   0.65  &   0.79  \\
\hline
$(\xi=3) \to (\xi=2)$ &   &      &       &       &       &       &       \\
& $0_{3,0}$ & $2_{2,1}$ & 400.00 & 257.90  & 205.27  & 178.52  & 123.22  \\
& $2_{3,1}$ & $4_{2,2}$ & 205.71 & 123.14  &  94.54  &  80.50  &  51.57  \\
& $2_{3,1}$ & $0_{2,0}$ &   0.00 &   0.22  &   0.38  &   0.46  &   0.54  \\
& $4_{3,2}$ & $6_{2,3}$ & 192.59 & 108.39  &  80.68  &  67.46  &  40.44  \\
& $4_{3,2}$ & $2_{2,1}$ &   0.00 &   0.34  &   0.58  &   0.69  &   0.79  \\
& $6_{3,3}$ & $8_{2,4}$ & 190.21 & 101.58  &  73.59  &  60.54  &  34.16  \\
& $6_{3,3}$ & $4_{2,2}$ &   0.00 &   0.42  &   0.71  &   0.84  &   0.92  \\
\hline
$(\xi=4) \to (\xi=3)$ &   &      &       &       &       &       &       \\
& $0_{4,0}$ & $2_{3,1}$ & 600.00 & 358.53  & 273.82  & 232.05  & 144.02  \\
& $2_{4,1}$ & $4_{3,2}$ & 308.57 & 173.79  & 129.12  & 107.67  &  62.88  \\
& $2_{4,1}$ & $0_{3,0}$ &   0.00 &   0.26  &   0.43  &   0.51  &   0.56  \\
& $4_{4,2}$ & $6_{3,3}$ & 288.89 & 154.60  & 112.08  &  92.13  &  50.93  \\
& $4_{4,2}$ & $2_{3,1}$ &   0.00 &   0.41  &   0.66  &   0.77  &   0.81  \\
& $6_{4,3}$ & $8_{3,4}$ & 285.31 & 145.99  & 103.53  &  84.01  &  44.16  \\
& $6_{4,3}$ & $4_{3,2}$ &   0.00 &   0.51  &   0.82  &   0.94  &   0.96  \\
\hline
\end{tabular}
\end{table}

\newpage

\begin{table}

\caption{Experimental spectra of the ground state bands of $^{100}$Pd 
\cite{Pd100} and $^{98}$Ru \cite{Ru98}, compared to the predictions
of the E(5)-$\beta^4$ and E(5)-$\beta^6$ models respectively.  
}

\bigskip

\begin{tabular}{r r r r r}
\hline
 L & $^{100}$Pd & E(5)-$\beta^4$ & $^{98}$Ru & E(5)-$\beta^6$ \\
\hline
 &       &       &       &       \\
 2  & 1.000 & 1.000 & 1.000 & 1.000 \\
 4  & 2.128 & 2.093 & 2.142 & 2.135 \\
 6  & 3.290 & 3.265 & 3.406 & 3.391 \\
 8  & 4.489 & 4.508 & 4.792 & 4.757 \\
 10 & 5.814 & 5.813 & 6.091 & 6.225 \\
 12 & 7.154 & 7.176 &       & 7.788 \\
 14 & 8.574 & 8.592 &       & 9.442 \\
 16 &10.425 &10.057 &       &11.180 \\
\hline
\end{tabular}
\end{table}

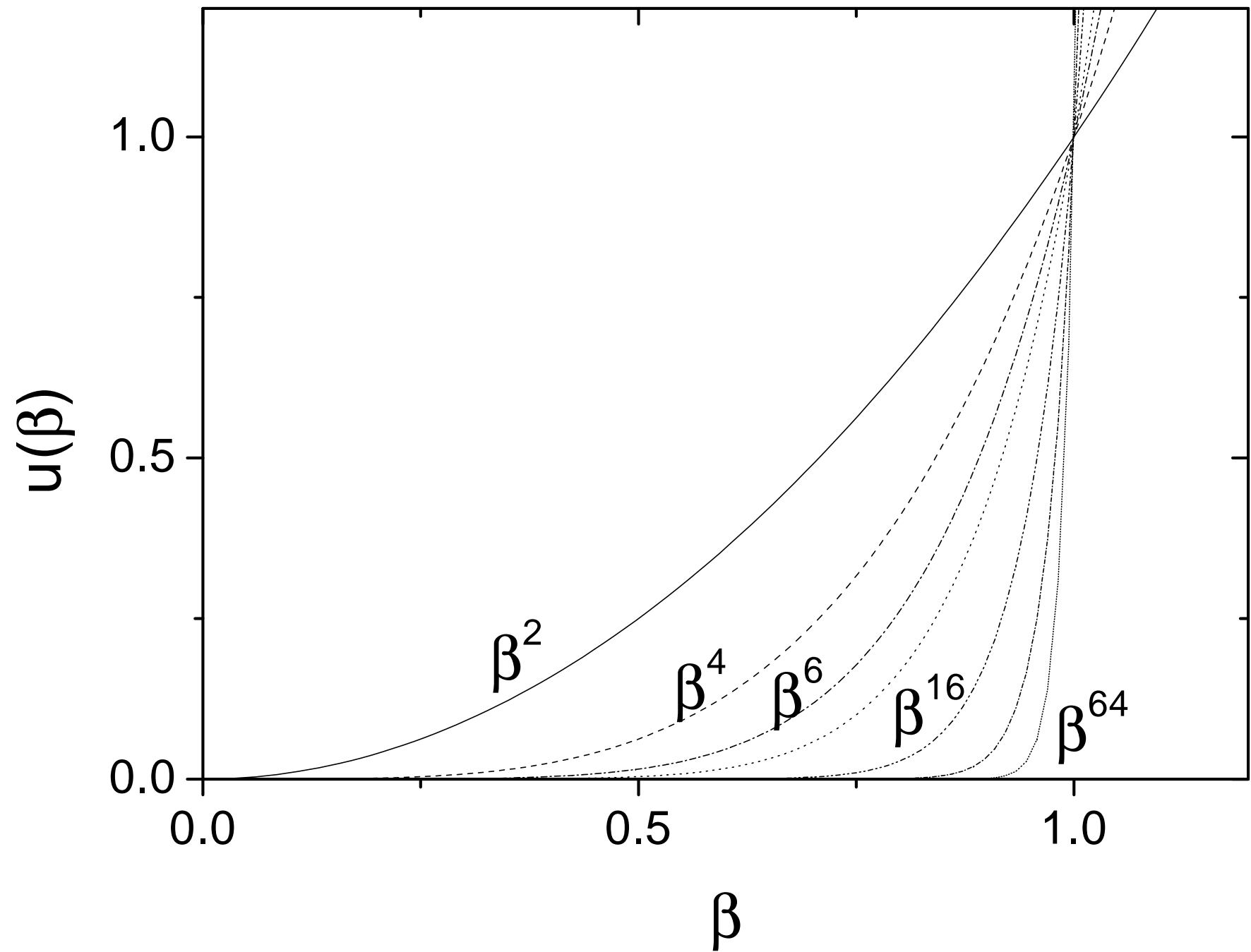
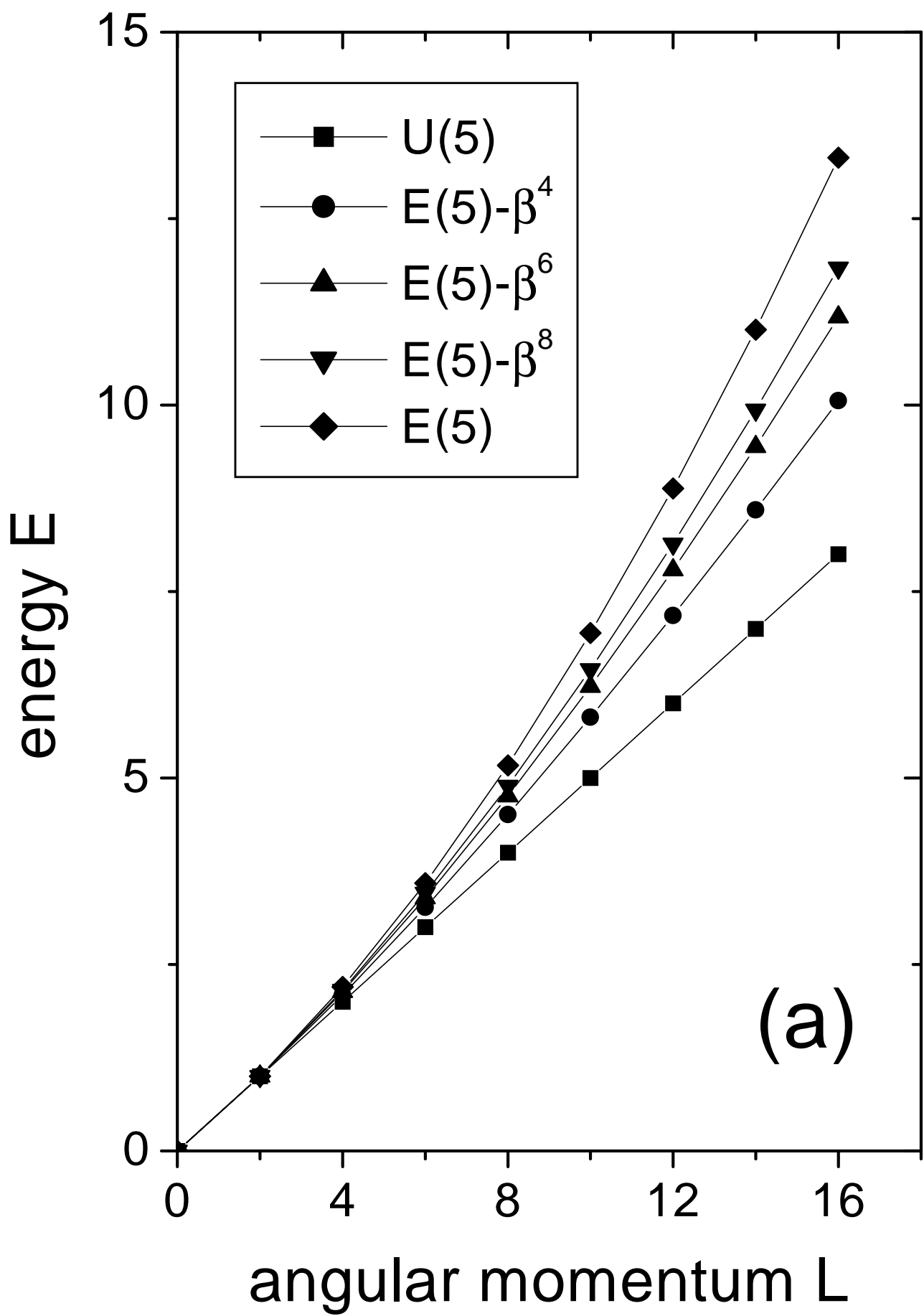
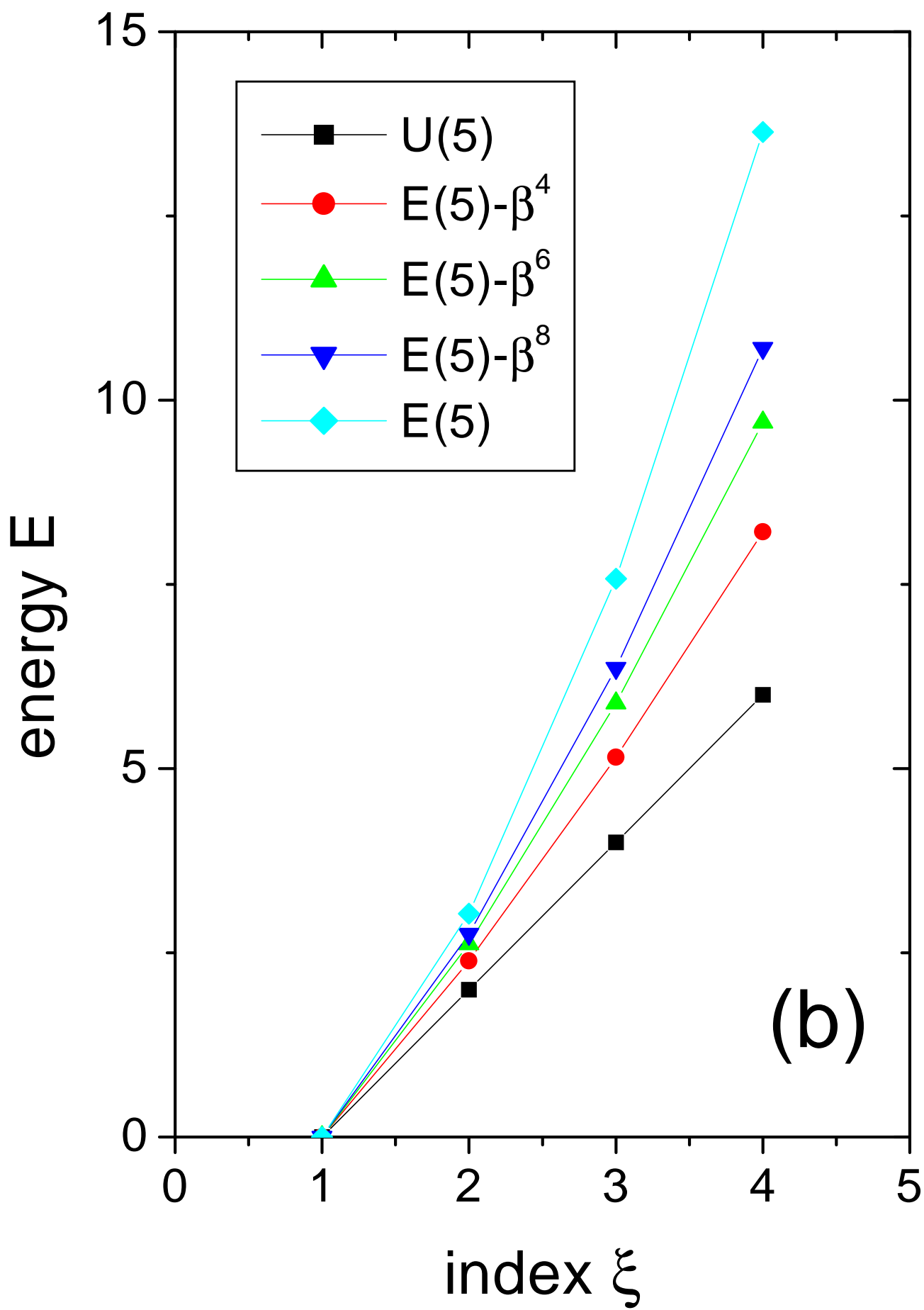
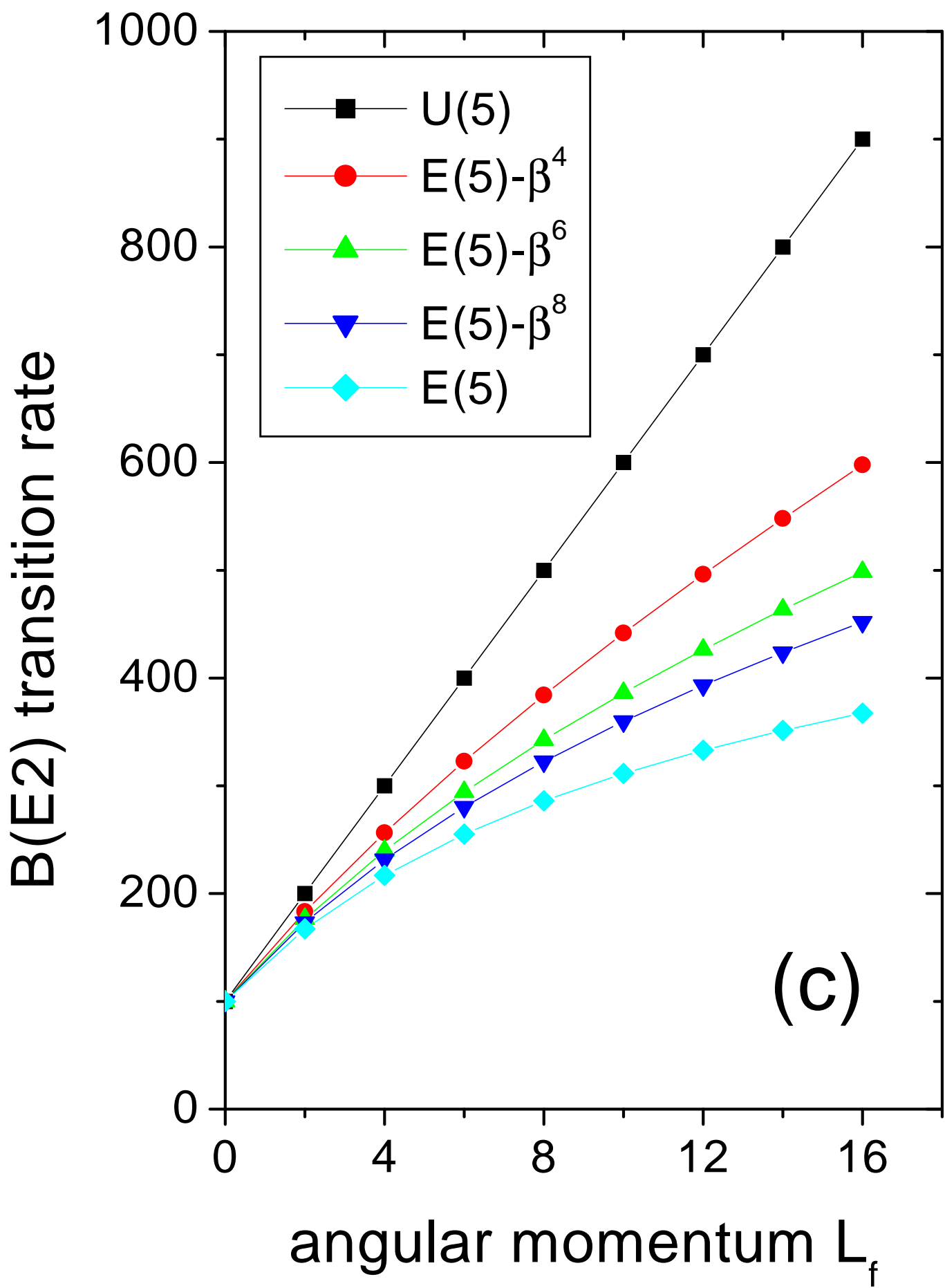
\end{document}